\documentclass[a4paper,11pt]{article}
\usepackage{pos}
\usepackage{xcolor}

\title{Calibration and Physics with ARA Station 1: A Unique Askaryan Radio Array Detector}
\ShortTitle{Calibration and Physics with ARA Station 1}

\author*[a]{Mohammad Ful Hossain Seikh}
\author[a]{Dave Besson}

\affiliation[a]{University of Kansas, Department of Physics and Astronomy, Lawrence, KS 66045, USA}

\onbehalf{for the ARA collaboration} 

\emailAdd{fulhossain@ku.edu}
\emailAdd{zedlam@ku.edu}

\abstract{The Askaryan Radio Array Station 1 (A1), the first among five autonomous stations deployed for the ARA experiment at the South Pole, is a unique ultra-high energy neutrino (UHEN) detector based on the Askaryan effect that uses Antarctic ice as the detector medium. Its 16 radio antennas (distributed across 4 strings, each with 2 Vertically Polarized (VPol), 2 Horizontally Polarized (HPol) receivers), and 2 strings of transmitting antennas (calibration pulsers, CPs), each with 1 VPol and 1 HPol channel, are deployed at depths less than 100 m within the shallow firn zone of the 2.8 km thick South Pole (SP) ice. We apply different methods to calibrate its Ice Ray Sampler second generation (IRS2) chip for timing offset and ADC-to-Voltage conversion factors using a known continuous wave input signal to the digitizer, and achieve a precision of sub-nanoseconds. We achieve better calibration for odd, compared to even samples, and also find that the HPols under-perform relative to the VPol channels. Our timing calibrated data is subsequently used to calibrate the ADC-to-Voltage conversion as well as precise antenna locations, as a precursor to vertex reconstruction. The calibrated data will then be analyzed for UHEN signals in the final step of data compression. The ability of A1 to scan the firn region of SP ice sheet will contribute greatly towards a 5-station analysis and will inform the design of the planned IceCube Gen-2 radio array.}

\ConferenceLogo{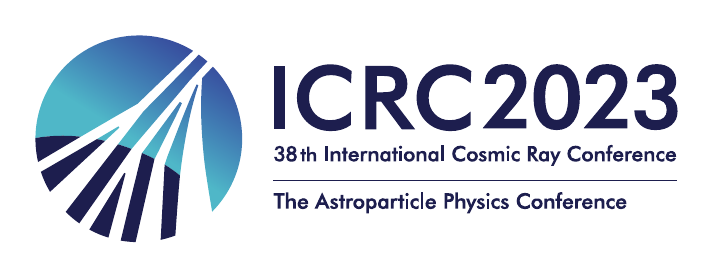}

\FullConference{%
38th International Cosmic Ray Conference (ICRC2023)\\
  26 July - 3 August, 2023\\
  Nagoya, Japan}

\begin{document}
\maketitle

\section{Introduction}
\begin{figure}[h]
		\centering\includegraphics[scale = 0.7]{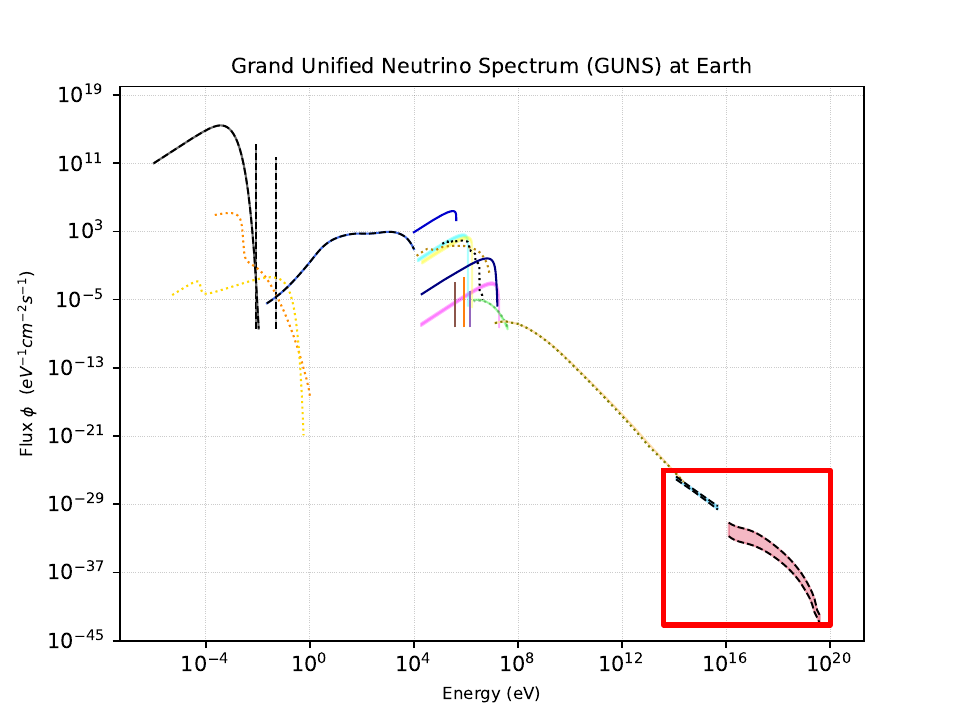}~
        \includegraphics[scale = 0.25]{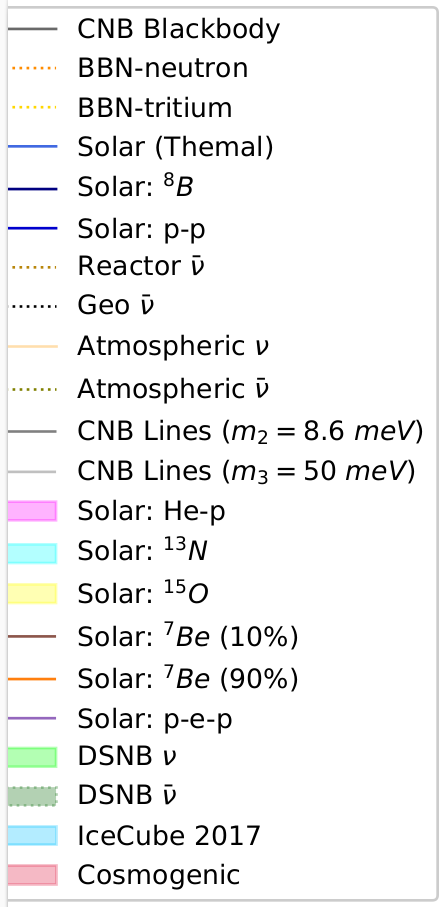}
		\caption{Neutrino spectrum at Earth integrated over directions and summed over flavors. Antineutrinos and  neutrinos are shown by dashed or dotted lines and solid lines, respectively. \textcolor{red}{Red} squared box illustrates the steep spectrum, particularly at TeV and EeV energies. Figure adapted from \cite{1}.}
		\label{fig:a}
\end{figure}

Neutrinos are unique cosmic messengers carrying important information about their sources over astronomical distance. Thus, detecting UHENs ($E_\nu \ge 10^{17}$ eV) opens a new window to the high energy universe. Whereas other UHE particles (namely cosmic rays and gamma rays) are either destroyed by interactions with the Cosmic Microwave Background (CMB) above approximately $10^{19.5}$ eV or annihilate with the CMB and/or Extragalactic Background Light above a few TeV, UHE neutrinos are largely unattenuated over cosmic distances due to their weakly interacting nature. However, the flux of UHEN is miniscule, as shown in Fig. \ref{fig:a}. Owing to their low flux and low cross-sections, we need gigantic detector volumes, for which we use naturally available ice sheets in Antarctica (ARA \cite{2}) and/or Greenland (RNO-G \cite{3}).
\subsection{Cosmogenic and astrophysical neutrinos}
When an accelerated proton from astrophysical objects such as Active Galactic Nuclei (AGN) or Gamma Ray Bursts (GRBs) interacts with the ambient matter (N), it produces pions which subsequently decay to neutrinos and gamma rays as shown in Eqns. \eqref{eq:1}, \eqref{eq:2} and \eqref{eq:3}.
\begin{align}
    p + N & \longrightarrow  \nonumber \pi + X~~~~(\pi = \pi^0, \pi^+, \pi^-)
     \\[1pt]
     \pi^0 & \longrightarrow \gamma + \gamma \label{eq:1}\tag{1}\\[1pt]
     \pi^+ (\pi^-) & \longrightarrow \mu^+ (\mu^-) + \nu_\mu (\bar\nu_\mu) \label{eq:2}\tag{2}\\[1pt]
    \mu^+ (\mu^-) & \longrightarrow e^+ (e^-) + \nu_e (\bar\nu_e) + \bar\nu_\mu (\nu_\mu)\label{eq:3}\tag{3}
\end{align}
The neutrinos produced in the environment host to these astrophysical objects are referred to as astrophysical neutrinos. Cosmogenic neutrinos ($E_\nu \ge$ 100 PeV), on the other hand, are produced when UHE cosmic rays (UHECRs) interact with the CMB as shown in Eqn. \eqref{eq:4}. 
\begin{align}
    p + \gamma_{CMB}  \xrightarrow[]{\hspace{2mm}\Delta^+\hspace{2mm}} 
    \begin{cases}
     p + \pi^0 \longrightarrow p + \gamma + \gamma\\
     n + \pi^+ \longrightarrow n + e^+ + \nu_\mu + \nu_e + \bar\nu_\mu    \end{cases}  
\label{eq:4}\tag{4}    
\end{align}
A source $\nu_e:\nu_\mu:\nu_\tau=1:2:0$ ratio evolves to $1:1:1$ at Earth via the phenomenon of neutrino oscillations. Detecting these UHENs will not only help us understand their sources but also will contribute significantly towards multi-messenger astronomy.

\section{The A1 Detector}
\label{sec:2}
A1 was deployed during the summer season of 2011-2012 and started consistently taking data in early 2014. Located 2 km away from the IceCube Neutrino Observatory, this station is unique in its configuration. Due to drilling difficulties, the in-ice antennas of this station could only be installed at a depth of less than 100 m instead of 200 m, such that the antennas are mostly in the shallow firn zone of the 2.8 km thick SP ice.
\begin{figure}[h]
        \centering\includegraphics[scale = 0.28]{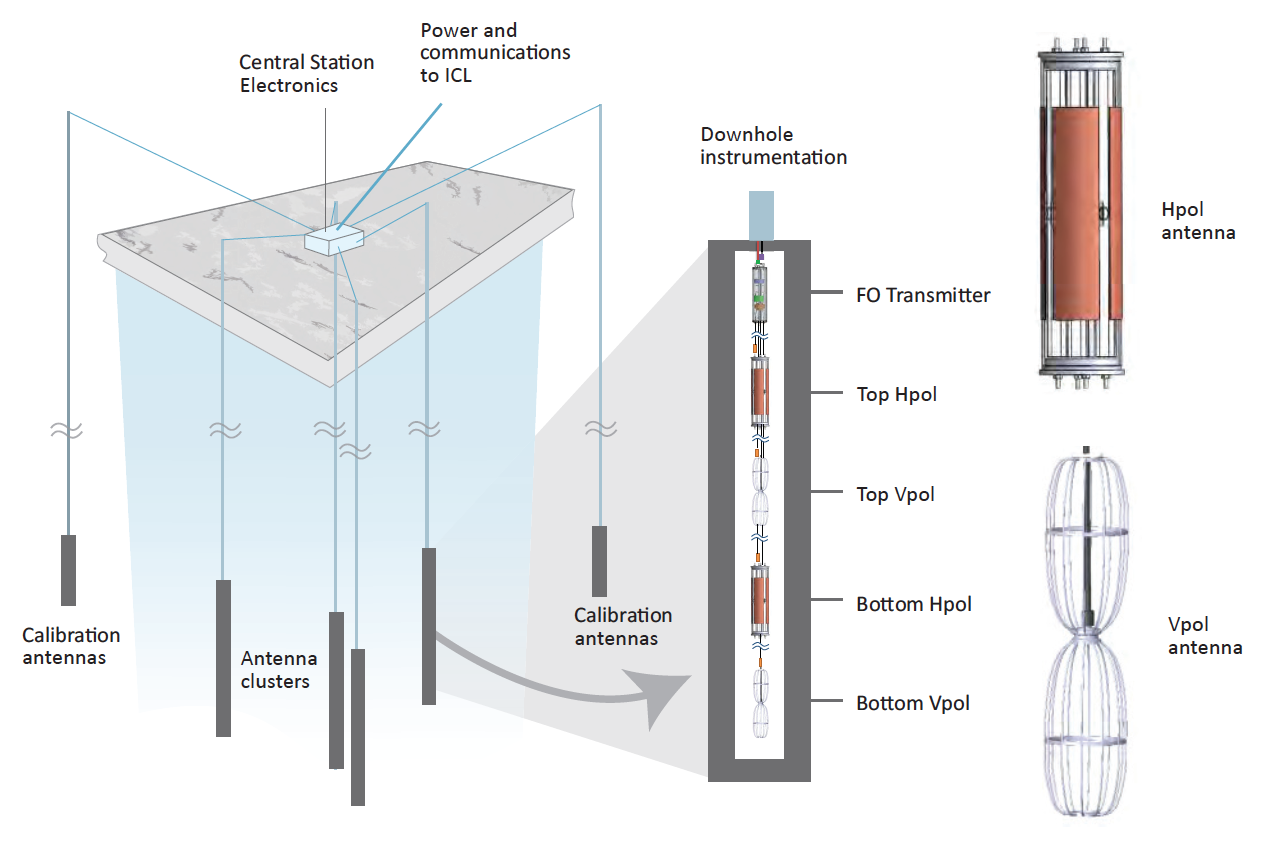}
		\caption{Schematic diagram of ARA station 1}
		\label{fig:b}
\end{figure}
 A schematic diagram of A1 with its transmitting and receiving antennas is shown in Fig. \ref{fig:b}. These antennas operate at radio frequencies, ranging from 150 MHz to 850 MHz. Four surface antennas offer cosmic ray detection. Signals from receivers are first notch filtered (at 450 MHz) to exclude narrow-band communication signals. Before being transmitted through low attenuation optical fiber to the surface, the filtered signals are amplified by Low Noise Amplifiers which contribute to a total signal chain gain of roughly 80 dB. Additional filters bandpass to 150-850 MHz signals at the surface DAQ. The signal then splits, and enters the Triggering Daughter board for ARA and the Digitizing Daughter board for ARA (DDA) which are mounted on a mother board (ARA Triggering and Readout Interface, or ATRI). The ATRI board also houses a Spartan-6 FPGA which contains the trigger logic. The digitizer records data only when a signal exceeds some pre-defined threshold in at least 3 of the 8 same-polarized antennas. The IRS2 chip then samples and digitizes with a sampling rate of up to 4 GHz (up to 8 channels simultaneously) and power consumption of less than 20 mW per channel. Each station has 4 IRS2 chips (DDA 0-3), each connected to 4 antenna channels. The high power efficiency and speed is typical of the Switched Capacitor Array architecture, comprising 128 capacitors and two delay lines per channel. Each delay line helps digitize 64 odd and 64 even samples in one odd-numbered and one even-numbered block; each block contains 64 samples. The analog data buffer can store 32k samples with a maximum buffer time of 10 $\mu s$, corresponding to a total of $32\times 1024 = 32768$ samples at a rate of $(32\times 1024)/(10\times 10^{-6}) \approx 3.2$ GS/s. A full readout therefore corresponds to $32768/64 = 512$ blocks of data. During sampling and digitization of these data, chip fabrication errors can induce a timing jitter of $\mathcal{O}$(100 ps), necessitating additional calibration of timing between consecutive samples \cite{4}.
 
\section{Timing Calibration}
Calibration is performed on lab data collected by plugging a continuous wave (CW) generator into the IRS2 chip at $-50^{\circ}~C$, similar to the SP deep ice temperature. This 214 MHz pre-deployment CW data had a bias voltage of $1.8~V$ which needs to be corrected (subtracted) before any calibration can be done. After pedestal correction, we estimate the difference between the recorded time and the expected time (jitter) of the IRS2 chip for each of the 128 sampling capacitors per channel. This can be achieved by the following steps:

\subsection{Separating odd-even samples}
A1 has 64 samples/block $\times$ 8 blocks = 512 samples in a waveform. For our calibration, we choose 384 samples (contrary to ARA stations 4 and 5 which used 896 samples \cite{5}) because the first block of most of the events is often corrupted and thus removed from calibration. Further filtering of corrupted events that starts with even blocks makes the timing array ({$\bf t_i$}, $i = 0, 1, ...., 383$) have an x-offset from zero, which is corrected for by subtracting the first element from each remaining element. As the odd and even samples lie on two different delay lines, we expect them to behave differently. We therefore fit sine waves separately to odd and even samples and check their fit frequencies ({$\bf f_i$}) for all the events, for all the channels. 

\subsection{Calibrating from the frequency distribution}
Now we histogram the frequency distribution of all the events, after eliminating outliers with the condition $\displaystyle{|f_i - \langle f_i\rangle| < 2\sigma(f_i)}$, and record the mean frequency ({$\bf f_m$}). We seek to converge this mean value to the input frequency ({$\bf f_I$ = 214~MHz}). We use another normalization condition $\displaystyle{t^\prime_i = \frac{f_m}{f_I}~t_i}$ to obtain a new timing array ({$\bf t^\prime_i$}).

\subsection{Minimizing jitter}
\begin{figure}[h]
    \begin{center}
\includegraphics[scale = 0.19]{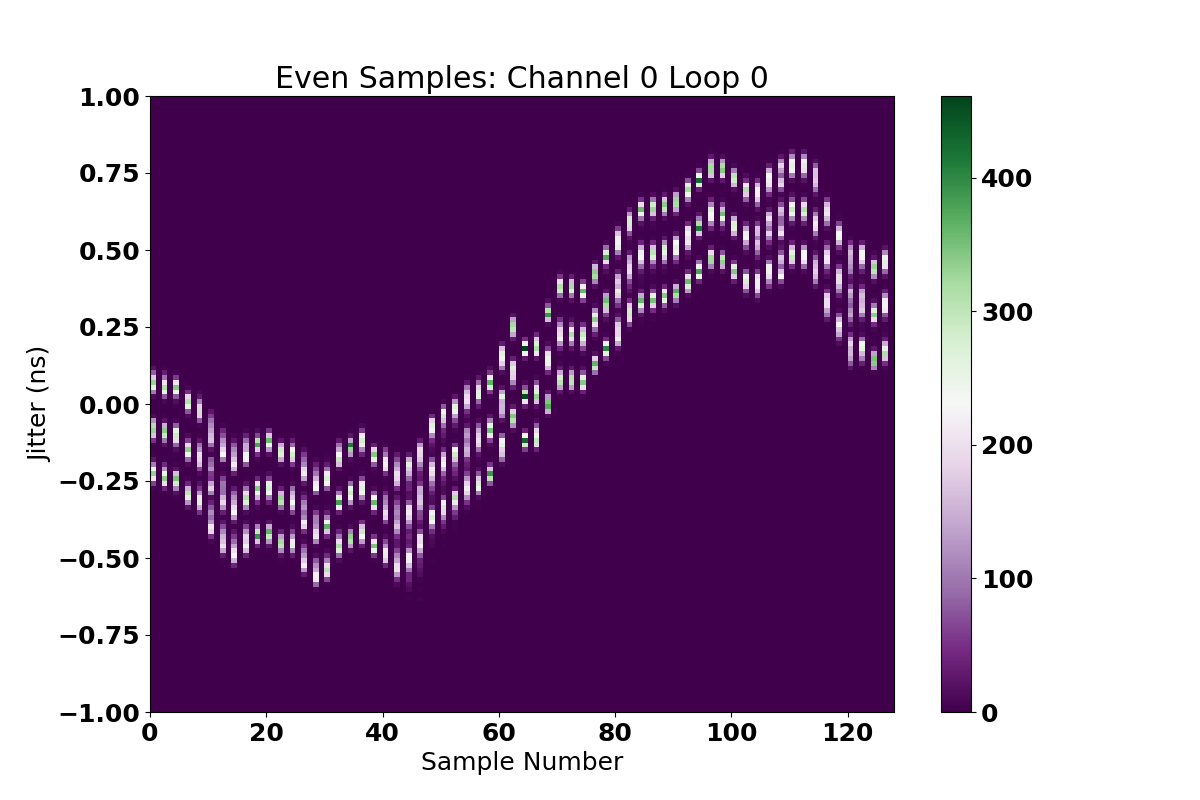}~\includegraphics[scale = 0.19]{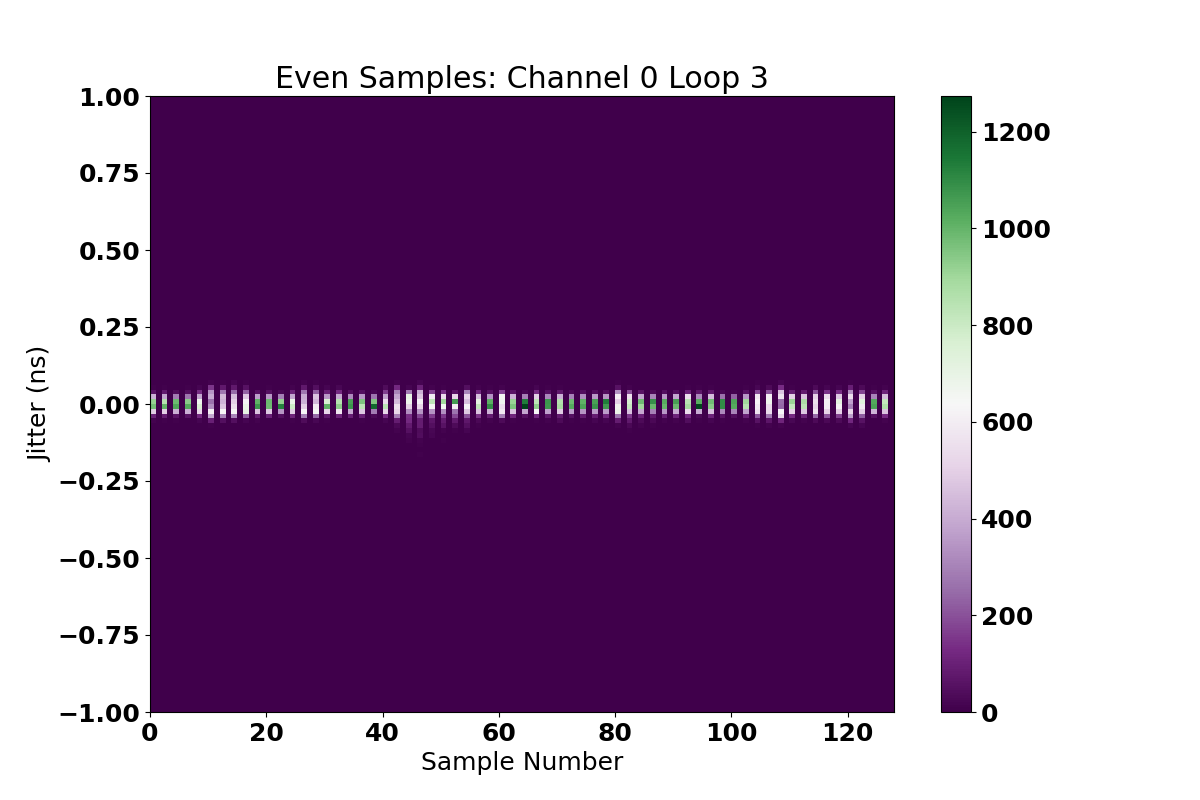}~\includegraphics[scale = 0.19]{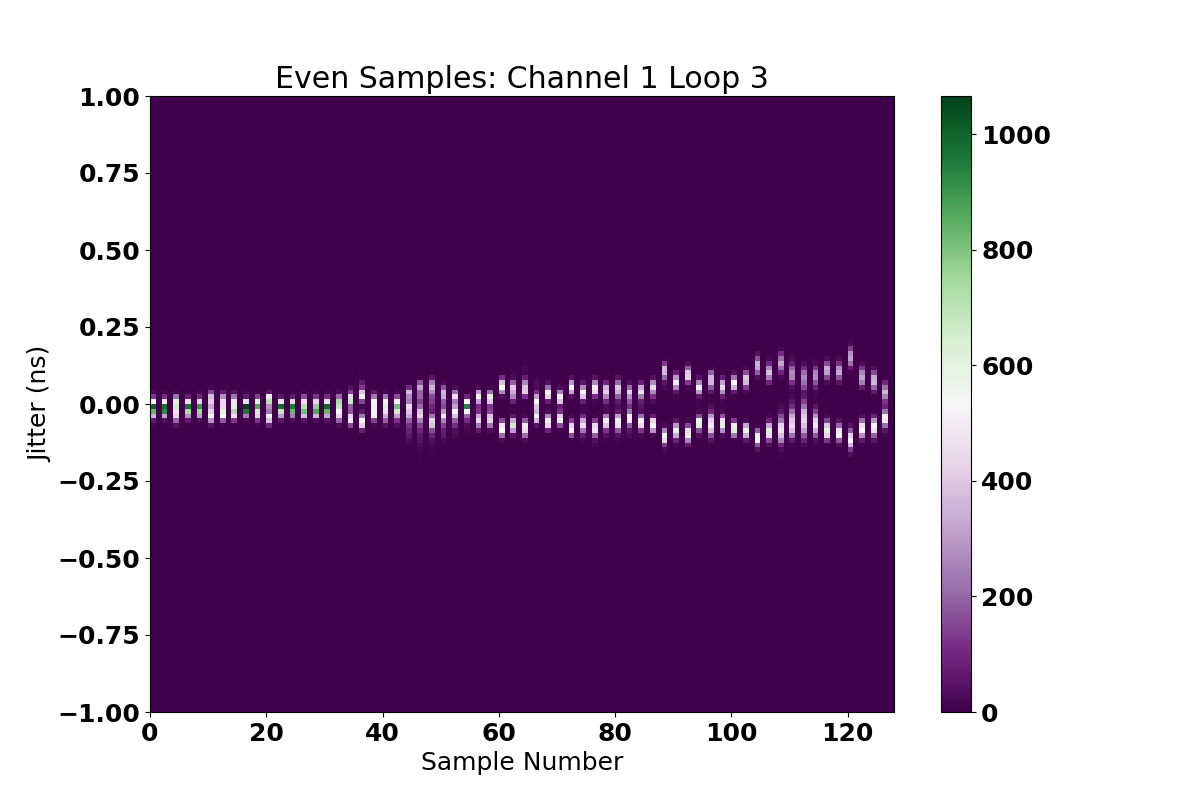}\\
\includegraphics[scale = 0.19]{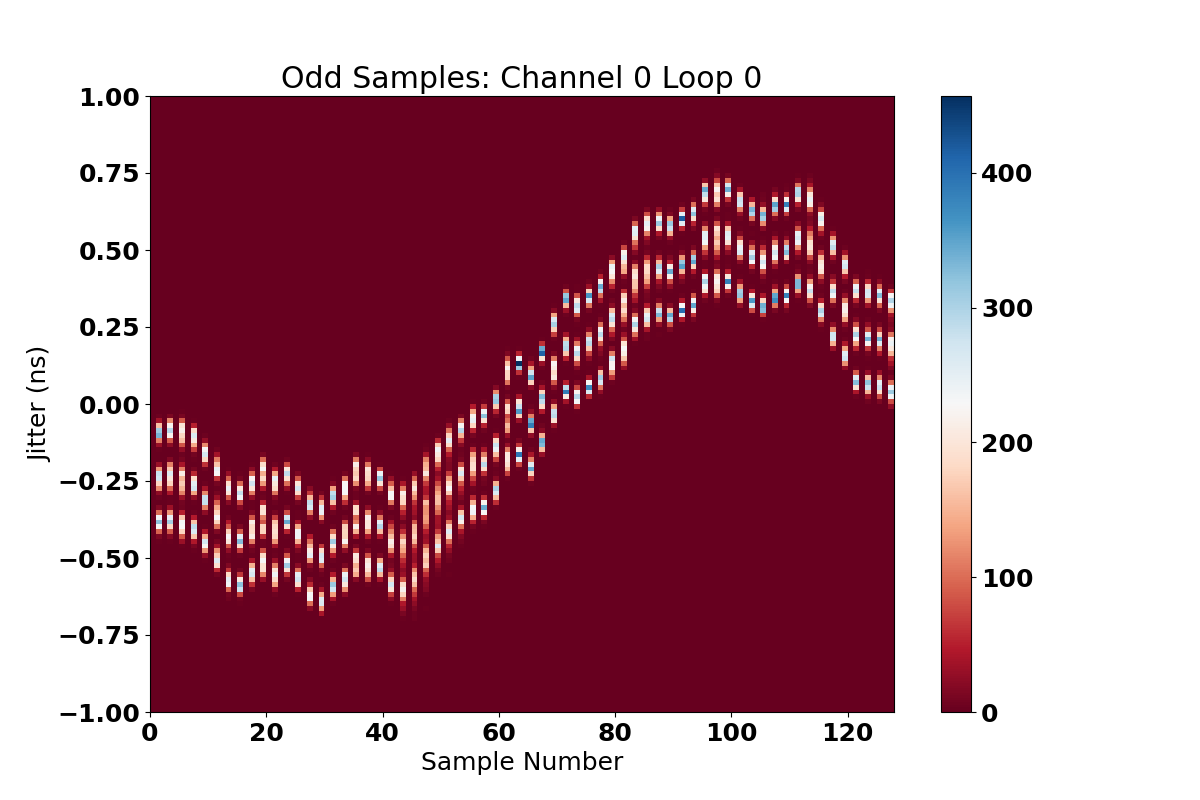}~\includegraphics[scale = 0.19]{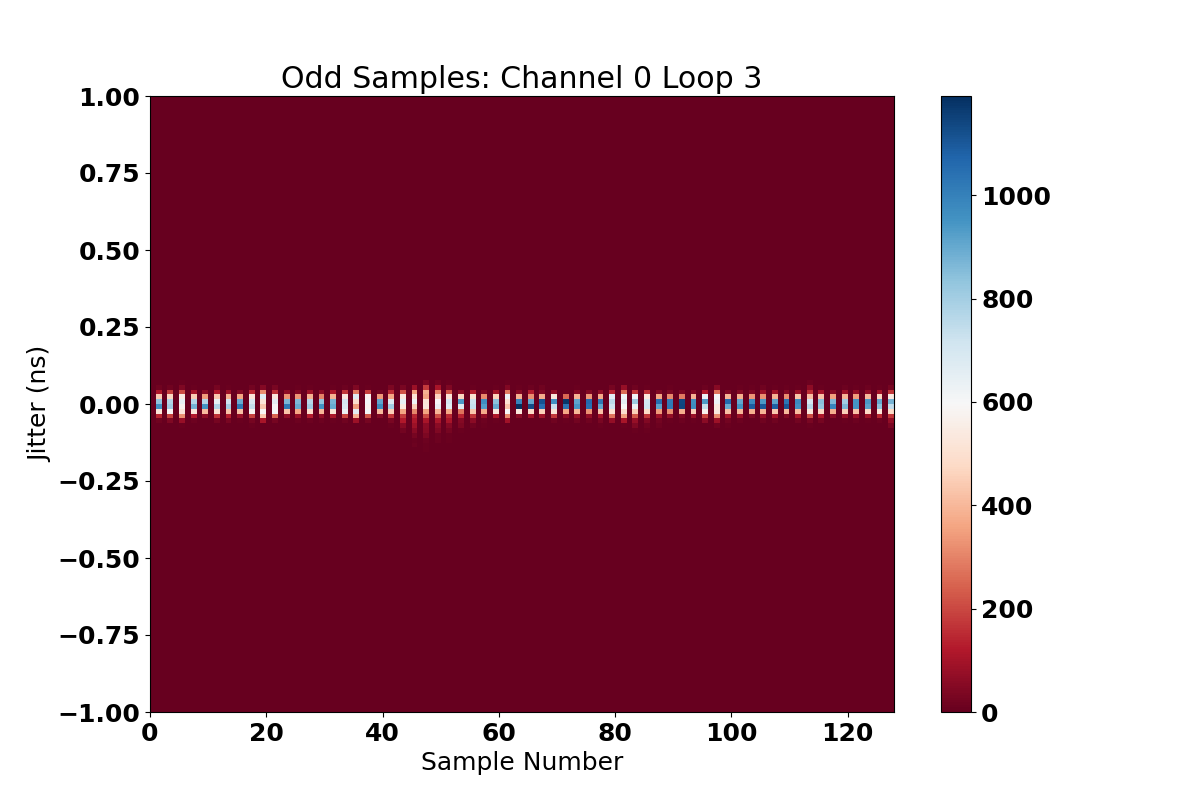}~\includegraphics[scale = 0.19]{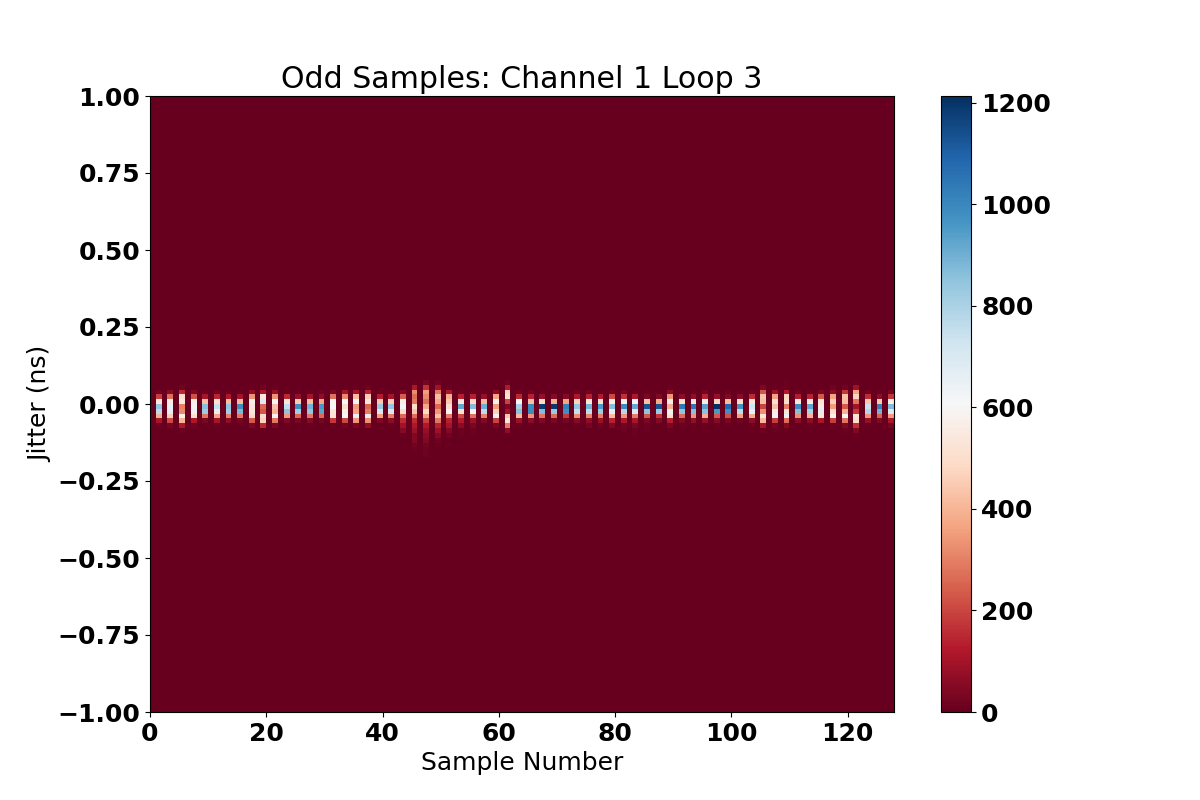}
    \end{center} 
    \caption{Jitter for even samples (top) and odd samples (bottom). The first column is pre-calibration, whereas the second is channel 0's post-timing resolution and the third is channel 1's post-timing resolution.}
    \label{fig:c}
\end{figure}

The jitter array ({$\bf j_i$}) is the timing offset between the expected and recorded times, for all of the 128 samples considered. To minimize this residual time, we repeat the above steps until the average jitter value for each sample approaches zero. This can be achieved within a few iterations; in each new iteration, the timing array is updated to:
$\displaystyle{t_i = t^\prime_i + j_i}$.\\\\
After calibration, the jitter vs.\ sample number distribution shows jitter centered at zero, as expected for a successful timing calibration. Even samples in channel 1 retain a bimodal behavior beyond calibration suggesting second-order corrections may be needed. We also notice that, in the same DDA, the performance of the chip degrades with increasing channel number, as can be see from the residual post-calibration jitter in the top right panel of Fig. \ref{fig:c}.

\section{Voltage Calibration}
The amplitude calibration of the input sine wave can be performed in two different ways. First, we can histogram the ADC counts for each sample to be calibrated. The peak positions are then used as the corresponding amplitude to fit a sine wave with running frequency and phase which are recorded as voltages. The ratio of ADC-to-voltage (ADC2V) is recorded for each sample as the ADC2V conversion factors are needed to convert our data into the desired voltage units. However, this method is a crude way of determining the ADC2V conversion factors as it is time consuming and not sufficiently precise, particularly when the conversion is not perfectly linear. Moreover, the  available calibration data is statistically insufficient for cross-check and validation of the above described method. Therefore, we use a second method which is based on the timing calibration, as follows:

\subsection{Sine wave fitting}
As discussed briefly in \nameref{sec:2} (Section 2), there are 32768 samples (64 samples distributed in each of 512 blocks) to be calibrated for each channel. Calibration run 975 has 36007 events each with a two-block-subtracted size of 384 samples, to give a total of $(36007\times 384)/32768 \simeq 422$ entries per samples available during the calibration process.
We first fit a sine wave of the form $\displaystyle{y_i = A_i\sin(2\pi k_i t- \phi_i)}$ to the already existing timing calibrated data with $k_i$ and $\phi_i$ as free parameters and $A_i$ fixed at 390, determined from the maximum amplitude of the timing calibrated ADC distribution. 

\subsection{Linear vs. cubic fit}
As a next step, we perform a fit to a scatter plot of voltage vs. ADC counts. Both linear and broken cubic fits (fitting negative and positive data separately) were performed; we find that A1 data fits best to a linear dependence, similar to A4 \cite{5}, but unlike A2, A3 \cite{6} and A5 \cite{5} which favored cubic correlations. In this step, we also discard the fit parameters for samples with reduced $\chi^2 \ge 1$ and choose the $\chi^2/ndf$ of the neighbouring sample for the same. 

\begin{figure}[h]
    \centering
    \includegraphics[height = 6cm, width = 8cm]{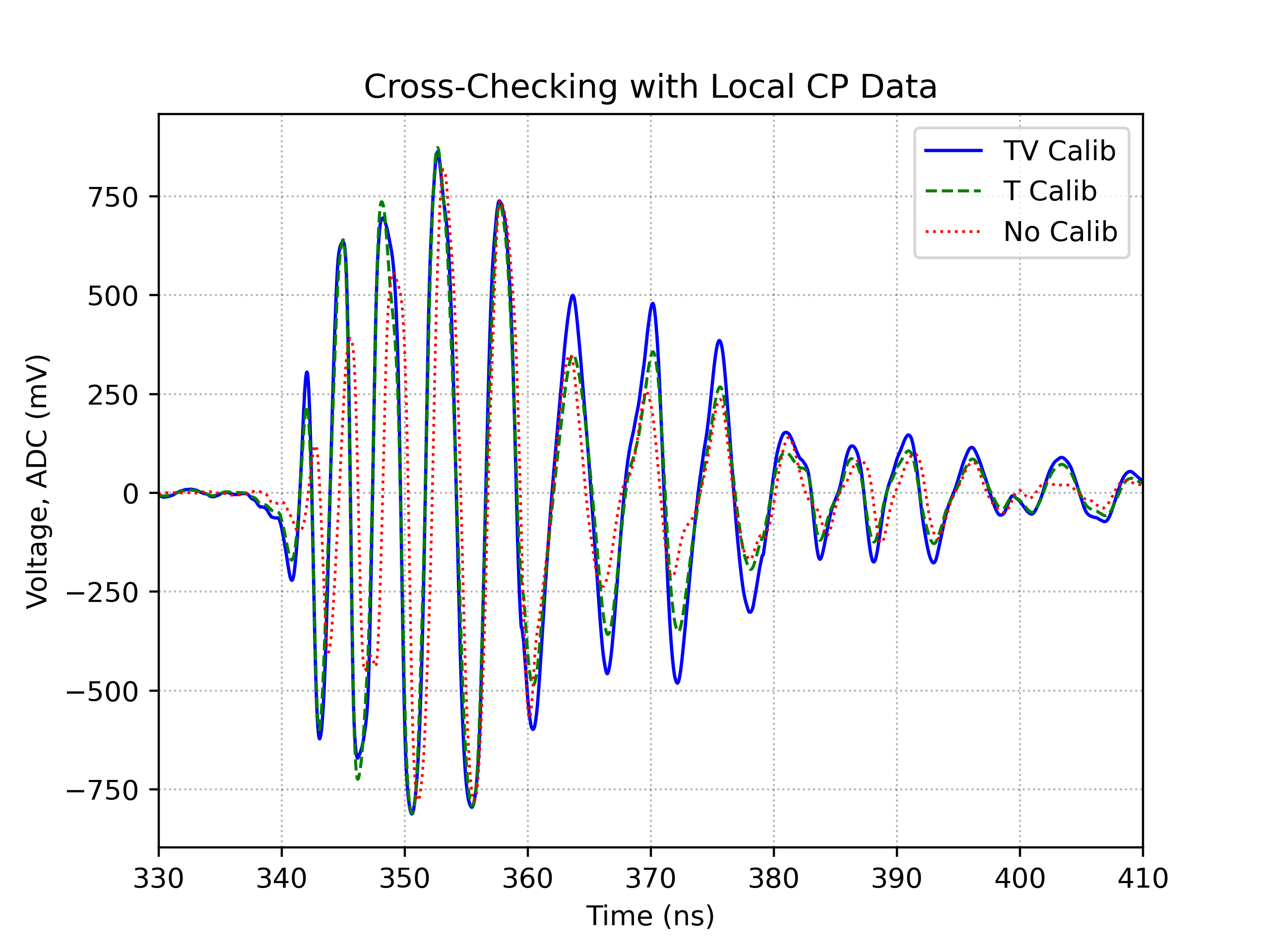}~
    \includegraphics[height = 6cm, width = 8cm]{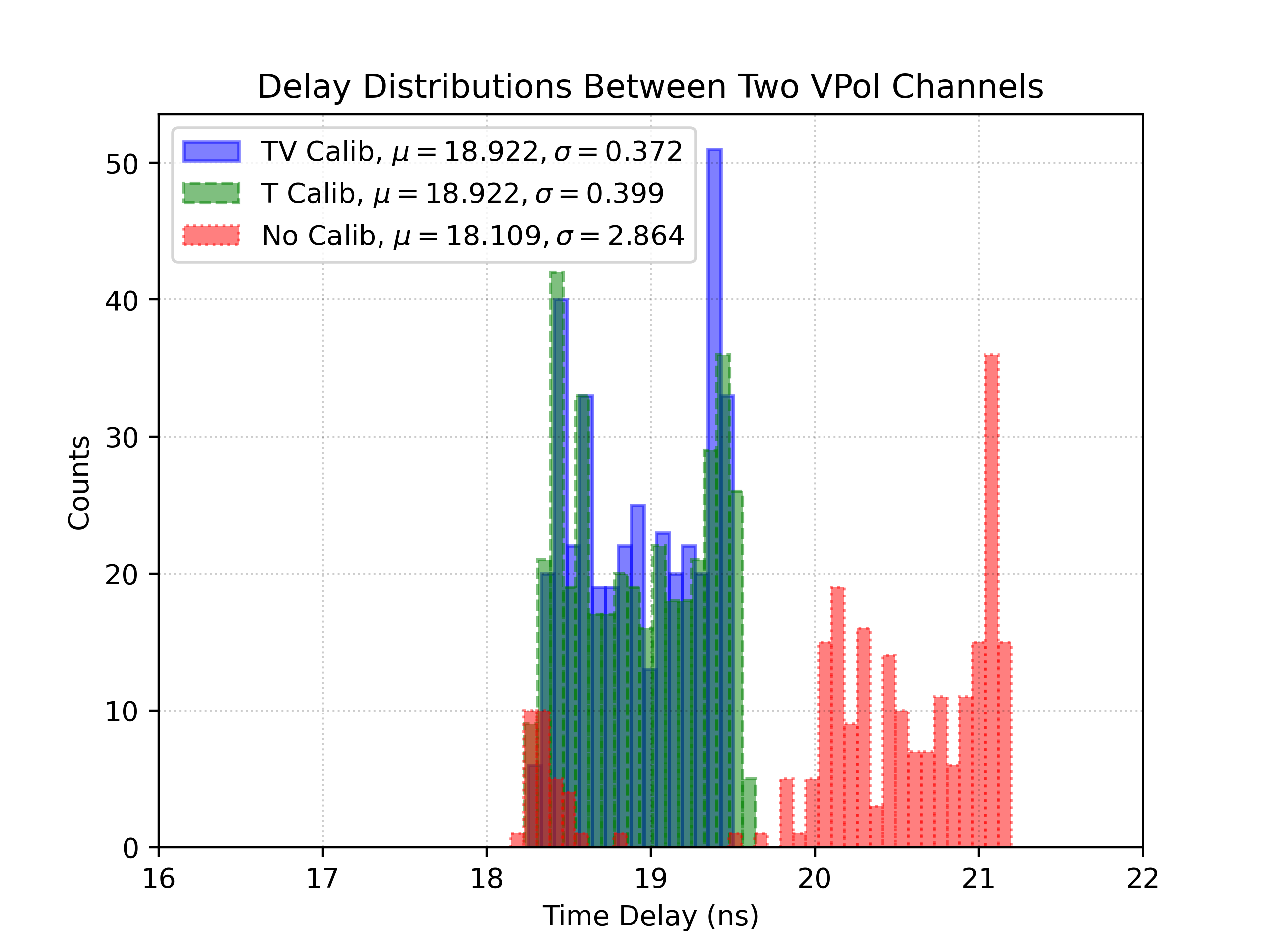}
    \caption{Application of calibration on CP data (left) and delay distributions between electric channel 9 and 16 (right).
    \textcolor{red}{Dotted-Red} color shows no calibration (No Calib), \textcolor{green!70!black}{Dashed-Green} color shows only timing calibration (T Calib), and \textcolor{blue}{Solid-Blue} color shows both timing and voltage calibration (TV Calib).}
    \label{fig:d}
\end{figure}
After completing the voltage calibration, we validate our result by considering real CP data. We choose a run, average all the CP events in a given run to improve the signal-to-noise ratio and note the average pulser time ($\bf \langle  T^{c1}_i \rangle $). Next, we find the time difference between the average waveform in a channel and all other events in that channel. Finally, for the signal arrival delay distribution between two channels, we take the difference $\displaystyle{\delta_{c1c2} = (\langle T^{c1}_i \rangle - T^{c1}_i) - (\langle  T^{c2}_i \rangle  - T^{c2}_i)}$ where $c1$ and $c2$ are the VPol channel numbers. A sample delay distribution between two VPol channels (and illustration of the application of calibration on CP waveforms) is shown in Fig. \ref{fig:d}. As evident from the standard deviation, the calibration narrows the pre-calibration delay distributions, after both timing and voltage calibrations. We also find slight improvement in the peak amplitude of the signal post-calibration.

\section{Antenna Position Calibration}
The in-ice antenna positions were surveyed during deployment with approximately half a meter accuracy but for a proper vertex reconstruction of neutrino source, we desire cm-scale precision. As a proxy for a neutrino source, we use known two calibration sources: the local CPs, two deep radio pulsers co-deployed with IceCube in 2011, and also the South Pole IceCore (SPIceCore) pulser \cite{7}. Optimizing antenna positions will also let us solidify cable delays and the ice model with respect to A1. The following steps were used for position calibration:

\subsection{Finding an ice model}
\begin{figure}[h]
    \centering
    \includegraphics[scale = 0.48]{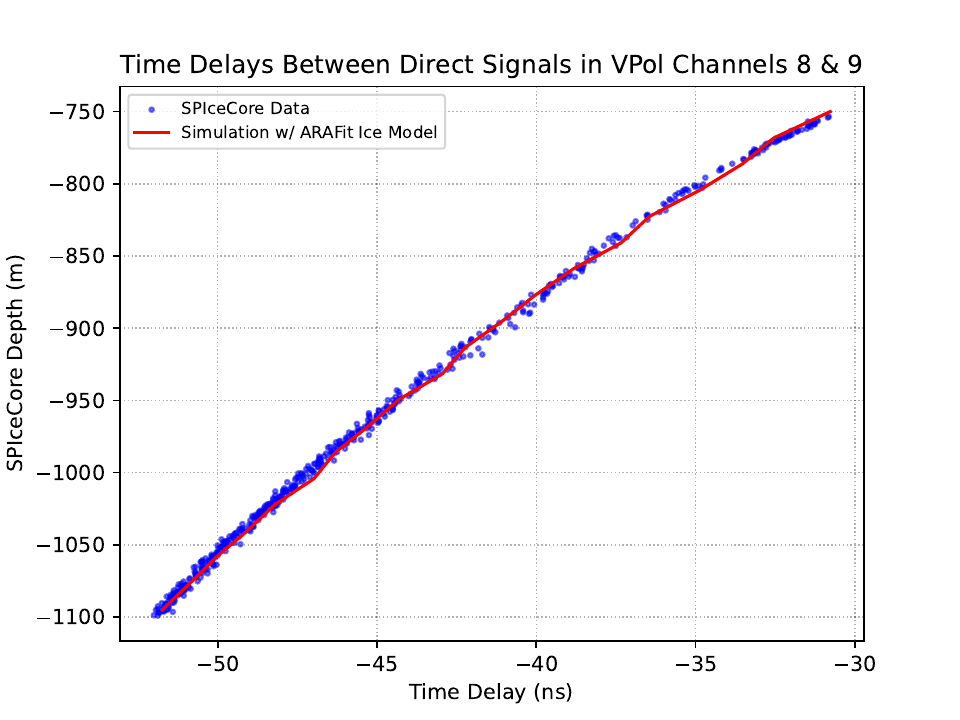}~
    \includegraphics[scale = 0.35]{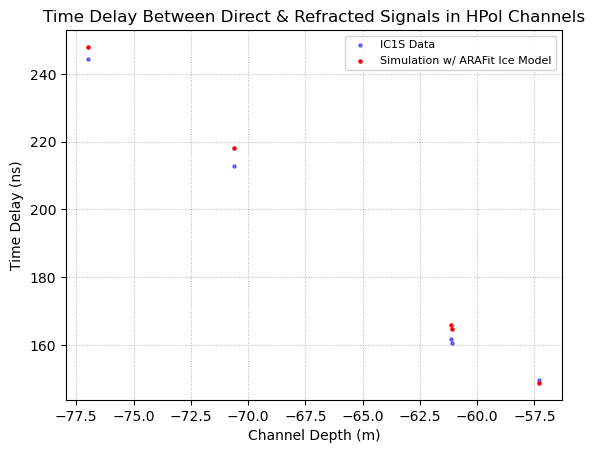}
    \caption{Time delays of SPIceCore (left) and IC1S (right).}
    \label{fig:e}
\end{figure}

We use SPIceCore data for which the transmitting antenna was dropped down from a depth of 850 m to 1100 m and find the relative time delays of direct signals between two VPol channels. Similarly, we find the time delays between direct and refracted pulses from deep pulsers at IceCube string 1 (IC1S) and string 22 (IC22S) for all HPol channels. We also use analytical ray-tracing and simulate the time delays for both the cases. As a final step, the data and simulation are fitted for the optimized depth ($z$) dependent refractive index model (ARAFit Ice Model),  $\displaystyle{n(z) = A - B e^{Cz}}$ with $A = 1.78$, $B = 0.42$, and $C = -0.0179$ as suggested by Fig. \ref{fig:e}.
\subsection{Finding optimized antenna locations}
A multi-parameter minimization is used to further calibrate the antenna geometry. An A1 specific ice model is used to calculate the ray-traced time ${\bf \Delta t_{RT}}$ between different channels for a variety of input data: channel-to-channel time delays from CP data (${\bf \Delta t_{CP}}$) and SPIceCore data as a function of depth of the transmitter source (${\bf \Delta t_{SP} (z)}$). These along with the cable delays data from the deployment ${\bf \Delta t_{CABLE}}$ are used in a Minuit optimizer to solve for the antenna locations, pulsers' (CP, SP) locations, and cable delays corresponding to the  minimum
 $\displaystyle{\chi^2_{final} = ({\bf \Delta t_{RT(CP)}} + {\bf \Delta t_{CP}} + {\bf \Delta t_{CABLE}})^2  + ({\bf \Delta t_{RT(SP)}(z)} + {\bf \Delta t_{SP} (z)} + {\bf \Delta t_{CABLE}})^2}$ value.

\section{Result and Conclusion}
\begin{figure}[h]
    \centering
    \includegraphics[scale = 0.4]{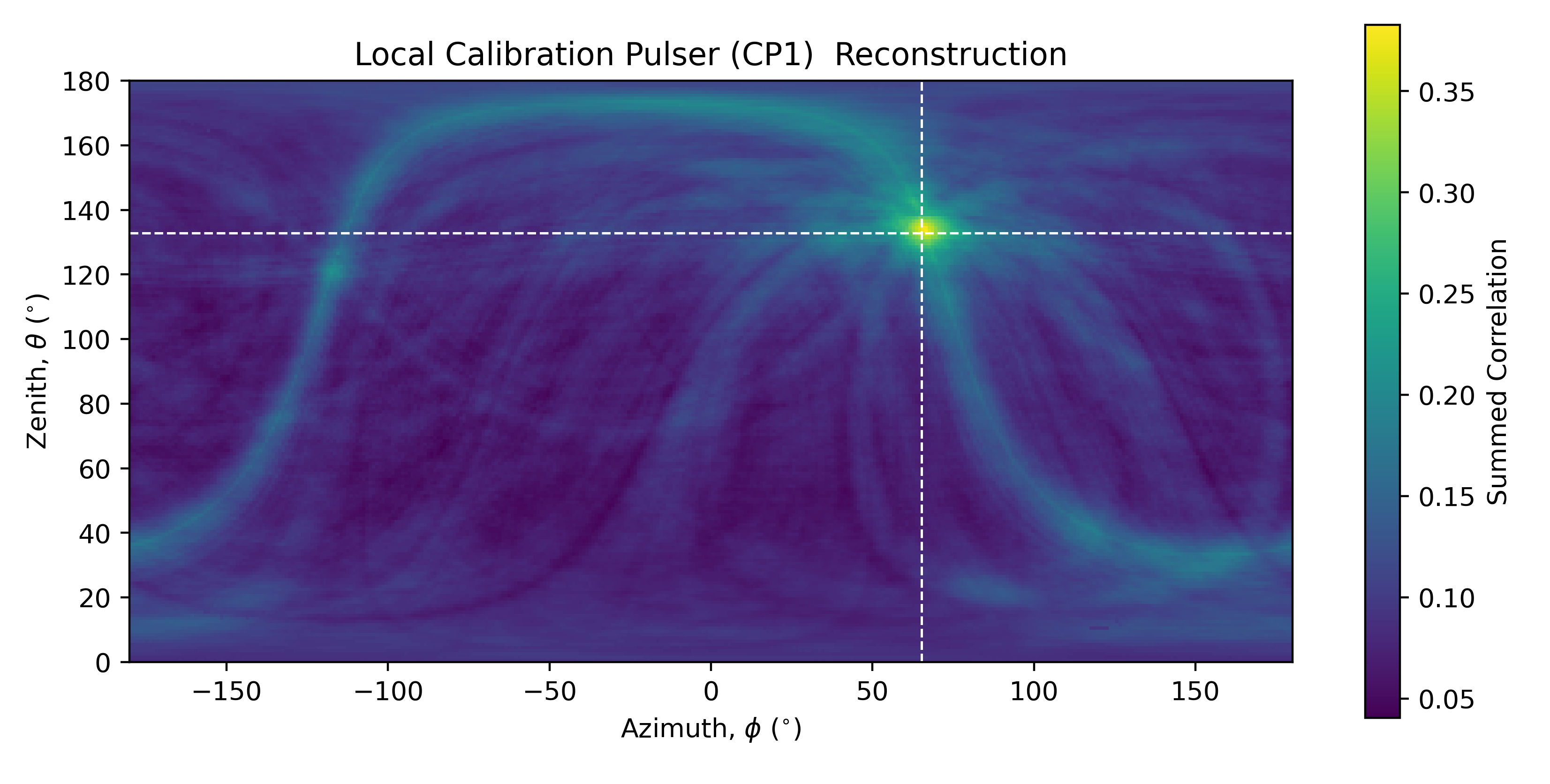}~
    \includegraphics[scale = 0.4]{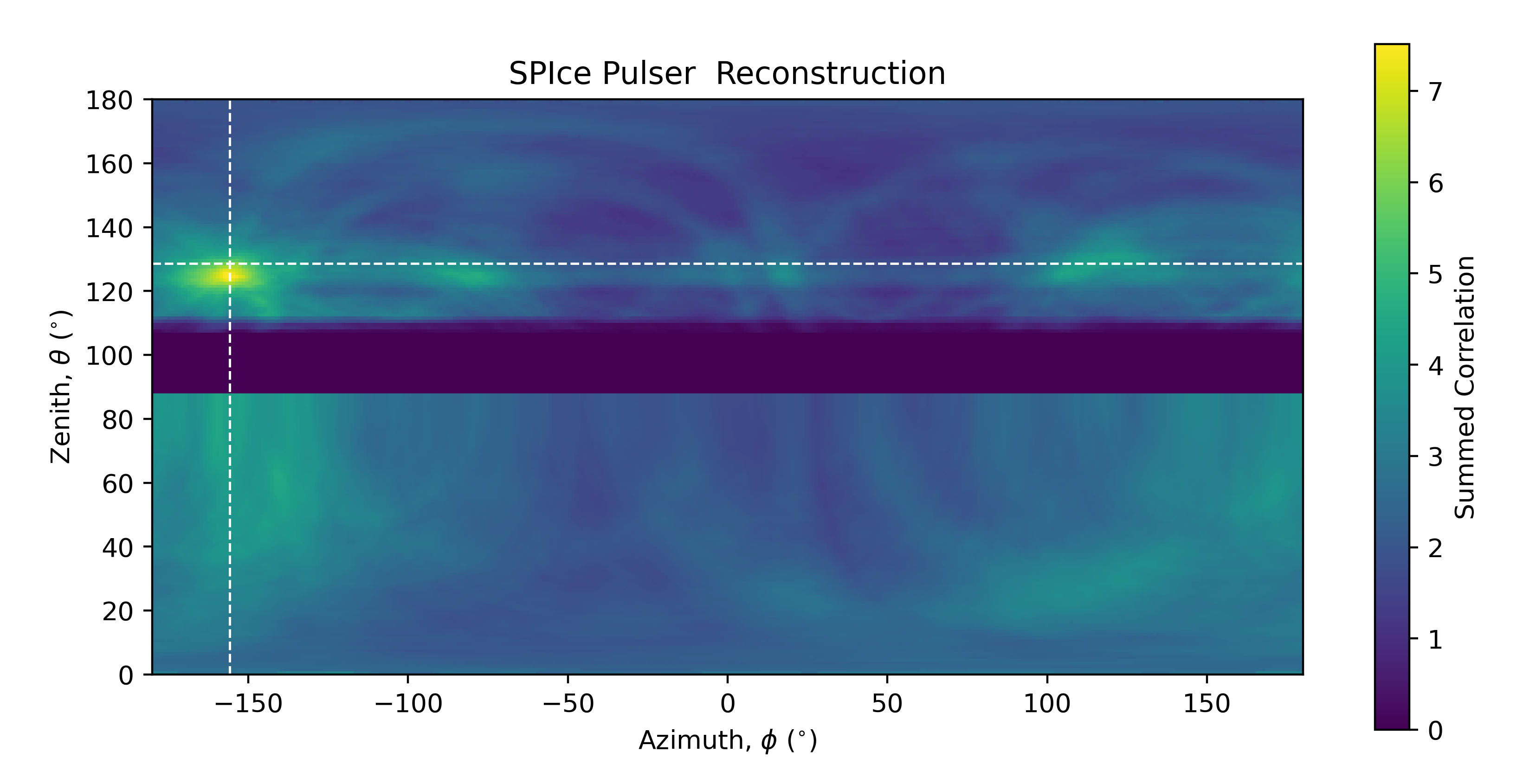}
    \caption{Reconstructed skymap of local calpulser (left) and SPIceCore pulser (right) at a depth z = -1066 m using calibrated A1 coordinates. Intersection of the White-Dashed lines is the actual vertex position.}
    \label{fig:h}
\end{figure}
As a final validation of our timing, voltage and antenna position calibration, we reconstruct our known local CP and distant SPIceCore pulser calibration sources (as shown in Fig. \ref{fig:h})  using the maximum summed correlation of ray traced and data hit time difference between different pairs of channels. We find a reasonable pointing resolution of zenith and azimuth angles with a few degrees of uncertainty. For the distant source, there is $\sim 2 ^{\circ}$ offset in the value of the reconstructed $\theta$ which is partly due to the error in the estimate of SPIceCore pulser depth as it was moving with time as well as bending of the direct ray by a few degrees in zenith depending on the depth and lateral distance of the source. The zero values in the summed correlation, which correspond to a \textcolor{blue!50!black}{dark blue} band in the right plot of Fig. \ref{fig:e} indicate a region where the ray-tracing solution does not exist. Overall, we achieve a precision in timing of $\mathcal{O}$(100 ps) and a statistical precision in antenna position of $\mathcal{O}$(10 cm) through calibration. A1 is now ready to analyze data in search of UHEN signals.

\clearpage
{
\section*{Full Author List: ARA Collaboration (July 18, 2023)}

\noindent
S.~Ali\textsuperscript{1},
P.~Allison\textsuperscript{2},
S.~Archambault\textsuperscript{3},
J.J.~Beatty\textsuperscript{2},
D.Z.~Besson\textsuperscript{1},
A.~Bishop\textsuperscript{4},
P.~Chen\textsuperscript{5},
Y.C.~Chen\textsuperscript{5},
B.A.~Clark\textsuperscript{6},
W.~Clay\textsuperscript{7},
A.~Connolly\textsuperscript{2},
K.~Couberly\textsuperscript{1},
L.~Cremonesi\textsuperscript{8},
A.~Cummings\textsuperscript{9}\textsuperscript{,}\textsuperscript{10}\textsuperscript{,}\textsuperscript{11},
P.~Dasgupta\textsuperscript{12},
R.~Debolt\textsuperscript{2},
S.~de~Kockere\textsuperscript{13},
K.D.~de~Vries\textsuperscript{13},
C.~Deaconu\textsuperscript{7},
M.~A.~DuVernois\textsuperscript{4},
J.~Flaherty\textsuperscript{2},
E.~Friedman\textsuperscript{6},
R.~Gaior\textsuperscript{3},
P.~Giri\textsuperscript{14},
J.~Hanson\textsuperscript{15},
N.~Harty\textsuperscript{16},
B.~Hendricks\textsuperscript{9}\textsuperscript{,}\textsuperscript{10},
K.D.~Hoffman\textsuperscript{6},
J.J.~Huang\textsuperscript{5},
M.-H.~Huang\textsuperscript{5}\textsuperscript{,}\textsuperscript{17},
K.~Hughes\textsuperscript{9}\textsuperscript{,}\textsuperscript{10}\textsuperscript{,}\textsuperscript{11},
A.~Ishihara\textsuperscript{3},
A.~Karle\textsuperscript{4},
J.L.~Kelley\textsuperscript{4},
K.-C.~Kim\textsuperscript{6},
M.-C.~Kim\textsuperscript{3},
I.~Kravchenko\textsuperscript{14},
R.~Krebs\textsuperscript{9}\textsuperscript{,}\textsuperscript{10},
C.Y.~Kuo\textsuperscript{5},
K.~Kurusu\textsuperscript{3},
U.A.~Latif\textsuperscript{13},
C.H.~Liu\textsuperscript{14},
T.C.~Liu\textsuperscript{5}\textsuperscript{,}\textsuperscript{18},
W.~Luszczak\textsuperscript{2},
K.~Mase\textsuperscript{3},
M.S.~Muzio\textsuperscript{9}\textsuperscript{,}\textsuperscript{10}\textsuperscript{,}\textsuperscript{11},
J.~Nam\textsuperscript{5},
R.J.~Nichol\textsuperscript{8},
A.~Novikov\textsuperscript{16},
A.~Nozdrina\textsuperscript{1},
E.~Oberla\textsuperscript{7},
Y.~Pan\textsuperscript{16},
C.~Pfendner\textsuperscript{19},
N.~Punsuebsay\textsuperscript{16},
J.~Roth\textsuperscript{16},
A.~Salcedo-Gomez\textsuperscript{2},
D.~Seckel\textsuperscript{16},
M.F.H.~Seikh\textsuperscript{1},
Y.-S.~Shiao\textsuperscript{5}\textsuperscript{,}\textsuperscript{20},
D.~Smith\textsuperscript{7},
S.~Toscano\textsuperscript{12},
J.~Torres\textsuperscript{2},
J.~Touart\textsuperscript{6},
N.~van~Eijndhoven\textsuperscript{13},
G.S.~Varner\textsuperscript{21},
A.~Vieregg\textsuperscript{7},
M.-Z.~Wang\textsuperscript{5},
S.-H.~Wang\textsuperscript{5},
S.A.~Wissel\textsuperscript{9}\textsuperscript{,}\textsuperscript{10}\textsuperscript{,}\textsuperscript{11},
C.~Xie\textsuperscript{8},
S.~Yoshida\textsuperscript{3},
R.~Young\textsuperscript{1}
\\
\\
\textsuperscript{1} Dept. of Physics and Astronomy, University of Kansas, Lawrence, KS 66045\\
\textsuperscript{2} Dept. of Physics, Center for Cosmology and AstroParticle Physics, The Ohio State University, Columbus, OH 43210\\
\textsuperscript{3} Dept. of Physics, Chiba University, Chiba, Japan\\
\textsuperscript{4} Dept. of Physics, University of Wisconsin-Madison, Madison,  WI 53706\\
\textsuperscript{5} Dept. of Physics, Grad. Inst. of Astrophys., Leung Center for Cosmology and Particle Astrophysics, National Taiwan University, Taipei, Taiwan\\
\textsuperscript{6} Dept. of Physics, University of Maryland, College Park, MD 20742\\
\textsuperscript{7} Dept. of Physics, Enrico Fermi Institue, Kavli Institute for Cosmological Physics, University of Chicago, Chicago, IL 60637\\
\textsuperscript{8} Dept. of Physics and Astronomy, University College London, London, United Kingdom\\
\textsuperscript{9} Center for Multi-Messenger Astrophysics, Institute for Gravitation and the Cosmos, Pennsylvania State University, University Park, PA 16802\\
\textsuperscript{10} Dept. of Physics, Pennsylvania State University, University Park, PA 16802\\
\textsuperscript{11} Dept. of Astronomy and Astrophysics, Pennsylvania State University, University Park, PA 16802\\
\textsuperscript{12} Universit\'{e} Libre de Bruxelles, Science Faculty CP230, B-1050 Brussels, Belgium\\
\textsuperscript{13} Vrije Universiteit Brussel, Brussels, Belgium\\
\textsuperscript{14} Dept. of Physics and Astronomy, University of Nebraska, Lincoln, Nebraska 68588\\
\textsuperscript{15} Dept. Physics and Astronomy, Whittier College, Whittier, CA 90602\\
\textsuperscript{16} Dept. of Physics, University of Delaware, Newark, DE 19716\\
\textsuperscript{17} Dept. of Energy Engineering, National United University, Miaoli, Taiwan\\
\textsuperscript{18} Dept. of Applied Physics, National Pingtung University, Pingtung City, Pingtung County 900393, Taiwan\\
\textsuperscript{19} Dept. of Physics and Astronomy, Denison University, Granville, Ohio 43023\\
\textsuperscript{20} National Nano Device Laboratories, Hsinchu 300, Taiwan\\
\textsuperscript{21} Dept. of Physics and Astronomy, University of Hawaii, Manoa, HI 96822\\
}
\section*{Acknowledgements}

\noindent
The ARA Collaboration is grateful to support from the National Science Foundation through Award 2013134.
The ARA Collaboration
designed, constructed, and now operates the ARA detectors. We would like to thank IceCube and specifically the winterovers for the support in operating the
detector. Data processing and calibration, Monte Carlo
simulations of the detector and of theoretical models
and data analyses were performed by a large number
of collaboration members, who also discussed and approved the scientific results presented here. We are
thankful to the Raytheon Polar Services Corporation,
Lockheed Martin, and the Antarctic Support Contractor
for field support and enabling our work on the harshest continent. We are thankful to the National Science Foundation (NSF) Office of Polar Programs and
Physics Division for funding support. We further thank
the Taiwan National Science Councils Vanguard Program NSC 92-2628-M-002-09 and the Belgian F.R.S.-
FNRS Grant 4.4508.01 and FWO. 
K. Hughes thanks the NSF for
support through the Graduate Research Fellowship Program Award DGE-1746045. B. A. Clark thanks the NSF
for support through the Astronomy and Astrophysics
Postdoctoral Fellowship under Award 1903885, as well
as the Institute for Cyber-Enabled Research at Michigan State University. A. Connolly thanks the NSF for
Award 1806923 and 2209588, and also acknowledges the Ohio Supercomputer Center. S. A. Wissel thanks the NSF for support through CAREER Award 2033500.
A. Vieregg thanks the Sloan Foundation and the Research Corporation for Science Advancement, the Research Computing Center and the Kavli Institute for Cosmological Physics at the University of Chicago for the resources they provided. R. Nichol thanks the Leverhulme
Trust for their support. K.D. de Vries is supported by
European Research Council under the European Unions
Horizon research and innovation program (grant agreement 763 No 805486). D. Besson, I. Kravchenko, and D. Seckel thank the NSF for support through the IceCube EPSCoR Initiative (Award ID 2019597). M.S. Muzio thanks the NSF for support through the MPS-ASCEND Postdoctoral Fellowship under Award 2138121.


\begin{thebibliography}{99}
\bibitem{1} \href{https://doi.org/10.1103/RevModPhys.92.045006}{Edoardo Vitagliano, Irene Tamborra, and Georg Raffelt Rev. Mod. Phys. 92, 045006 2020}
\bibitem{2} \href{https://arxiv.org/pdf/1404.5285.pdf}{P. Allison et al., Astroparticle Physics 35 (457) 2012}
\bibitem{3} \href{https://iopscience.iop.org/article/10.1088/1748-0221/16/03/P03025}{J.A. Aguilar et al., 2021 JINST 16 P03025}
\bibitem{4} \href{https://journals.aps.org/prd/abstract/10.1103/PhysRevD.93.082003}{P. Allison et al., Phys. Rev. D 93, 082003 (2016)} \href{https://arxiv.org/pdf/1507.08991.pdf}{arXiv:1507.08991v3}
\bibitem{5} \href{https://inspirehep.net/files/3ccb9bf64badcc2dab4820abc9b77f15}{Paramita Dasgupta and Kaeli Hughes (ARA), PoS(ICRC2021)1086}
\bibitem{6} \href{https://link.springer.com/book/10.1007/978-3-319-18756-3}{Thomas Meures, ISBN: 978-3-319-18756-3}
\bibitem{7} \href{https://iopscience.iop.org/article/10.1088/1475-7516/2020/12/009}{P. Allison et al JCAP12(2020)009}
\end{thebibliography}
\end{document}